# IDENTIFICATION AND MOLECULAR DYNAMIC SIMULATION OF FLAVONOIDS FROM MEDITERRANEAN SPECIES OF OREGANO AGAINST THE ZIKA NS2B-NS3 PROTEASE

**Anushikha Ghosh[1], Arka Sanyal[1] and Sameer Sharma[2*]**

[1]Department of Biotechnology, KIIT University.

[2*]Department of Bioinformatics, Bio Nome Pvt. Ltd.

## ABSTRACT

**Objective:** The Zika virus (ZIKV), is an emerging infectious disease causing severe complications such as microcephaly in infants and Guillain–Barré syndrome in adults. There is no licensed vaccination or approved medicine to treat ZIKV infection. Therefore, extensive research is being carried out to find compounds that can be used effectively as therapeutic molecules to treat ZIKV infection. Oregano, a commonly found herb in the Mediterranean region, has been used predominantly for culinary purposes. The fact that the members of the *Origanum* species are a storehouse of various bioactive compounds gives us a solid reason to study compounds extracted from it for therapeutic purposes. **Methods:** In this study, were retrieved 20 Flavonoids found in various *Origanum* species belonging to the Mediterranean region from the PubChem database and pharmacological analysis using SwissADME and Molecular docking using AutoDock Vina 4.0. were carried out against the NS2B-NS3 protease since it serves as an effective drug target owing to its role in viral replication and immune evasion within the host. The best hit compound(s) were subjected to MD simulation at 100 ns using Desmond Schrodinger to analyze the molecule's stability. **Results:** We observed *Cirsiliol* as the best hit compound against the NS2B-NS3 complex with a binding affinity of -8.5 kcal/mol. It also showed good stability during MD simulation. **Conclusion:** We recommend the use of *Cirsiliol* for in vitro and in vivo studies for further investigation concerning the ZIKA virus.

## INTRODUCTION

Zika Virus (ZIKV) is an emerging mosquito-borne virus belonging to the Flaviviridae family and Flavivirus genus.[1] According to the current studies, there are 53 virus species belonging to this genus that are transmitted mainly via mosquito bites, although ticks and other unknown arthropod vectors are also associated with the transmission of this disease.[2] The ZIKV shows close phylogenetic relations with other mosquito-borne flaviviruses of, such as the Japanese encephalitis virus (JEV), dengue virus (DENV), yellow fever virus (YFV), West Nile virus (WNV).[1] ZIKV has spread expeditiously to 86 countries and territories, becoming a worldwide health issue.[4] Due to the increasing number of ZIKV outbreaks worldwide, it was declared a public health emergency by the World health organization on February 1, 2016.[5]

The ZIKV is an enveloped virus that is spherical and has an icosahedral-like symmetry.[6] The genome of ZIKV, as shown in Fig.1, comprises a positive-sense single-stranded RNA molecule approximately 11 kb (10.8 kb) long. The RNA molecule is divided into three regions - a 5' Untranslated Region (UTR), about 100 nucleotides long, a single Open Reading Frame (ORF) about 10 kb long, and a 3' UTR, about 420 nucleotides long. The ORF is responsible for coding a single polyprotein precursor of 3423 amino acids and processing them into three structural and seven non-structural proteins.[7] The structural proteins are in charge of the virus particle formation and are associated with the virus entry, assembly, and release of newly formed virions into the host cell.

On the other hand, the non-structural proteins play an essential role in the viral replication and packaging of the genome, thus destabilizing the host pathways and eventually favoring the virus.[8] As per studies, two critical non-structural proteins are associated with viral replication within the host. They are (a) Non-Structural protein 3 (NS3), which essentially comprises a serine protease at its N-terminal, while at its C-terminal, it comprises RNA triphosphatase and RNA helicase; (b) Non-Structural protein 2B (NS2B) which interacts and complexes with the NS3 protein to form the serine protease, essential for the viral activity.[9] The two-protein complex NS2B-NS3 protease aids in developing antivirals against ZIKV by acting as an excellent drug target. This protein complex plays a vital role in the replication of the virus within the host cell apart from helping the virus from invading the innate immune system.[10]

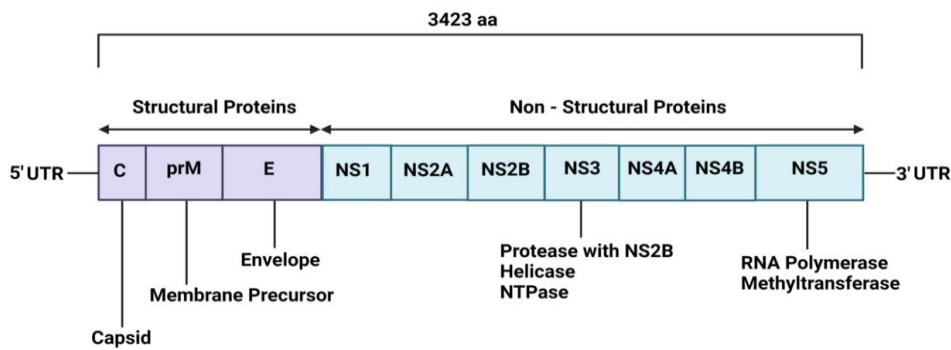

**Fig. 1: The figure shows a detailed structure of the Zika Virus (ZIKV) genome showing the viral RNA, which encodes for a single polyprotein precursor of 3423 amino acids cleaved co-translationally to yield ten proteins. Three of these are structural (C, E, and PrM) proteins, while the remaining seven are non-structural (NS1, NS2A, NS2B, NS3, NS4A, NS4B, and NS5) proteins.**

Currently, there are no preventive medicines or vaccines available to combat ZIKV. Since ZIKV infection usually appears asymptomatically or with mild symptoms, those may be controlled with bed rest, intravenous fluids, and paracetamol.[11] The most common signs and symptoms of ZIKV infection typically occur within a week after being bitten by a mosquito carrying the virus, which includes headache, fatigue, fever, rash, conjunctivitis, arthralgia, and myalgia.[12] Recent outbreaks of the disease have revealed that the virus is also capable of causing multi-organ failure[13], thrombocytopenia[14], meningitis, and encephalitis.[15] The major challenge of the Zika virus is the infection complication and its various transmission modes, especially maternal-fetal transmission that causes microcephaly[16] and Guillain–Barré syndrome.[17] The recent outbreak of ZIKV in Indian subcontinents indicates the fast spreading of the virus in the South-Eastern regions of Asia, which may lead to further outbreaks in the future.[18] To date, suitable treatment options, vaccines, and drugs are yet to arrive on the market to combat ZIKV. Therefore the only effective way to tackle the situation is its prevention. The general preventive measures include preventing mosquito breeding, especially in epidemic areas, and sexual transmission. Our study explores several phytochemicals as an alternative to find novel antiviral drugs against ZIKV.

Flavonoids refer to a class of phytochemicals that naturally occur in most plants as secondary polyphenolic compounds and are associated with various biological processes. Flavonoids are widely distributed across the members of the plant kingdom and are most commonly found in fruits, vegetables, and seeds. Flavonoids are generally depicted as C6-C3-C6 compounds and have a basic structure that consists of a fifteen-carbon skeleton.[19] Flavonoids have various pharmacological properties which draw the attention of researchers. These mainly include antioxidant properties, anti-cancerous, anti-inflammatory, antimicrobial, and antiviral functions.[20] Flavonoids undoubtedly are the most studied group of polyphenols and are persistently being researched by scientists to explore and discover their antiviral activities against different viruses.[21]

There are very few reported studies on Molecular Docking studies against Zika Virus.[22-25] To our knowledge, molecular docking studies using the flavonoids found in *Origanum* species are yet to be reported. Here, we report a novel in-silico study using flavonoids that have been reported to be found in Mediterranean species of Oregano. In our present study, we have focused explicitly on the *Origanum* species [*O. Dictamus, O. majoram* L. , and O.*vulgare* L.] found in the Mediterranean region [Greece, Tunisia, Italy, and Turkey] and accordingly selected 20 flavonoids [*Apigenin, Arbutin, Aromadendrin, Catechin, Chrysin, Cirsiliol, Cirsimaritin, Desmethoxycentauridin, Diosmetin, Eriodictyol, Genestein, Isovitexin, Kaempherol, Luteolin, Quercetin, Sakuranetin, Salvigenin, Sorbifolin, Taxifolin,* and *Xanthomicrol*] widely distributed among these species.[26-31] Subsequently, we retrieved their structures from the PubChem database, and subsequently performed molecular docking and molecular dynamic studies with these flavonoids, intending to find potential flavonoids possessing antiviral activity against our protein target, the ZIKV NS2B-NS3 Protease.

**MATERIALS AND METHODS**

**Ligand Preparation**

In this study, we shortlisted 20 flavonoids from various Origanum species found in the Mediterranean regions and extracted their 3D from the PubChem database.[32] All the 3D structures of the ligands were downloaded in the form of Standard Data Format (SDF) and later converted into the Protein Data bank (PDB) format with the help of Open Babel (version 3.1.1).

**Retrieval of Receptor**

NS2B-NS3 protease, a suitable drug target molecule, was used as a receptor against the shortlisted phytocompounds to inhibit ZIKV entry and further replication within the host body. The 3D structure of the receptor was downloaded from the Protein Data Bank directly.[33]

**Active Binding Site prediction**

The binding pockets of the receptor were determined using the online software Computed Atlas of Surface Topography of proteins (CASTp 3.0).[34] CASTp web servers provide a comprehensive computable characterization of protein pockets on the surface of the protein, thereby helping study the geometric and topological properties of protein structures.

**ADMET Analysis**

SWISS ADME[35], a drug screening tool, was used to analyze the compounds based on their ADMET (Chemical absorption, distribution, metabolism, excretion, and toxicity). We used several parameters like physicochemical, pharmacological, and drug-likeness to screen the phytocompounds. The drug-likeness of the selected phytocompounds was examined and identified using these parameters. All the physicochemical properties were strictly scrutinized based on Lipinski's filter[36], Ghose's filter[37], Veber's filter[38], Egan's filter[39], and Muegge's filter[40], and Bioavailability score.[41] Any compound that did not satisfy these parameters was exempted from further experimentation due to its lack of drug-likeness property.

**BOILED-Egg**

The Brain Or IntestinaL EstimateD permeation method (BOILED- Egg)[42] is a SWISS ADME feature used as an accurate predictive model to analyze the gastrointestinal absorption, blood-brain barrier, and permeability glycoprotein substrate of a particular compound. The compounds with high GI absorption lie clearly in the white region, and compounds with a high blood-brain barrier lie in the yellow region. To analyze the BOILED-Egg model, we used SWISS ADME.

**Molecular Docking Analysis**

The primary purpose of Molecular docking is to examine the inhibitory activities of the selected phytocompounds against the NS2B-NS3 protease receptor through various interactions. We performed docking on all the 20 shortlisted compounds using the virtual screening software known as PyRx.[43] PyRx includes various software needed for computational drug discovery, such as AutoDock 4.2, AutoDock Vina, AutoDockTools, Open Babel, Python, wxPython, etc. It uses AutoDock Vina and AutoDock 4.2 to perform the docking of the ligands for the target receptor. The receptor and the ligands were loaded into the PyRx workspace and converted into AutoDock input files (pdbqt files). After converting them successfully, they were docked using AutoDock Vina. The grid box was set at X=48.3353, Y=59.2129, and Z=63.1111 to cover the entire protein-ligand complex since blind docking was performed. After successful docking, we saved the docking results showing the binding affinity (kcal/mol).

We then used the PatchDock web server to calculate the docking scores of the following 20 flavonoids for molecular docking.[44] The receptor protein and the ligands were loaded one by one, we set the cluster RMSD as 1.5, and results containing the docking scores were retrieved. Four molecules were then shortlisted based on their binding affinities and docking scores which were analyzed at Biovia Discovery Studio.[45] The structural analysis of these compounds was done on Biovia Discovery Studio Visualizer. The interactions of the ligands with the receptor were analyzed via the 2D and 3D images of the ligand-receptor complex.

**Molecular Dynamics (MD) Simulation**
We performed a Molecular Dynamics (MD) simulation study on the top two molecules (Isovitexin and Cirsiliol) with the most interactions with the protein target as viewed on Discovery Studio. Molecular Dynamics Simulation is a computer-based simulation approach used to analyze the physical motions of atoms or molecules. A few critical hydrogen bond interactions can be identified using MD simulations. MD simulations aid protein docking and virtual screening advances. The iMODS server was used to simulate molecular dynamics in this work. The iMODS service facilitates normal mode analysis exploration by generating essential data about routes that may involve macromolecules or homologous structures.

For the hit chemical receptor complex, molecular dynamics simulations were also run using the Desmond program. In terms of dynamic studies, we have used Schrodinger Desmond package 4.0, which analyzed the stability of the complex structure received from docking Studies. Desmond will consider the free energy calculation that will be based on the temperatures. Both the protein and ligand were prepared to employ an OPLS force field followed by Sodium (Na+) and calcium ions (Cl-) for neutral charges. The desmond algorithm was continued for 100 nanoseconds (ns) and subjected to an NPT ensemble. The data was examined in terms of protein and ligand RMSD and root means square fluctuation (RMSF).

**RESULTS**

**Ligand Preparation**

All the 3D structures of the 20 shortlisted ligands were extracted from the PubChem database in Standard Data Format (SDF) and then were converted into Protein Data Bank (PDB) format using Open Babel (version 3.1.1), as shown in Fig.2.

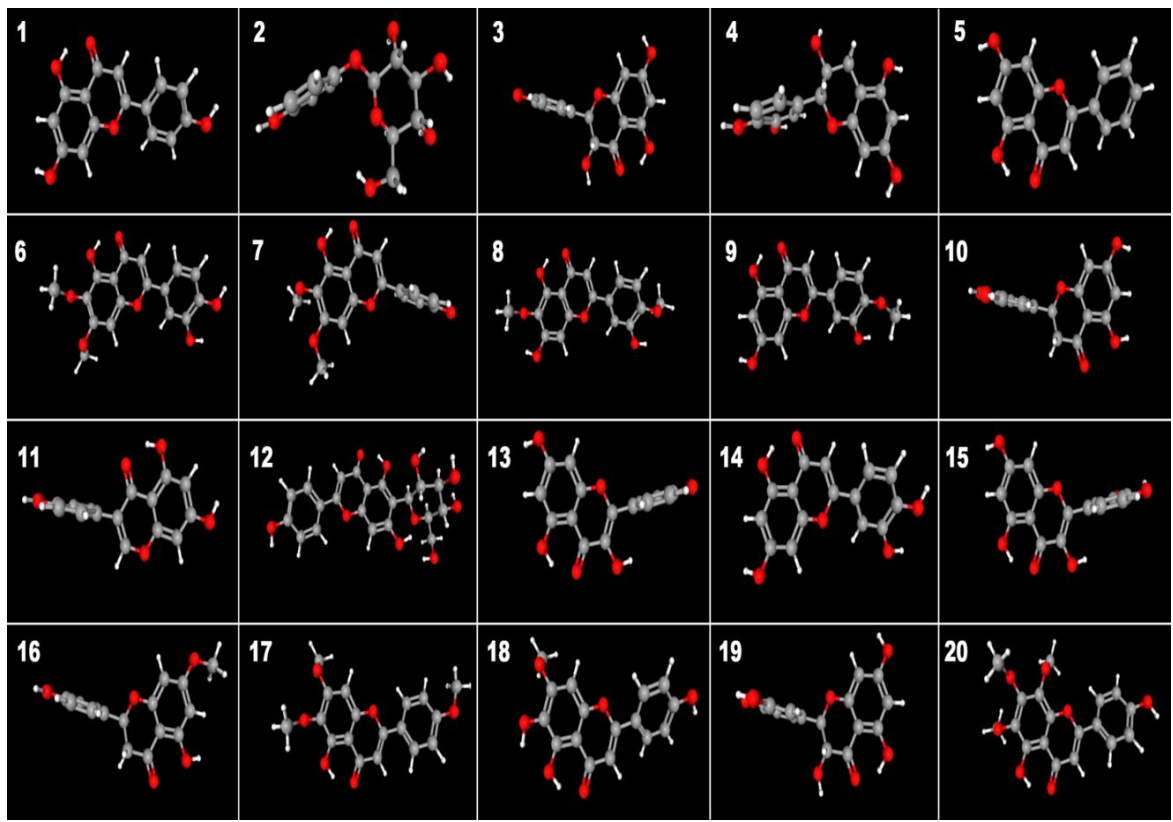

**Fig.2: The figure shows the 3D structures of the 20 compounds selected in our study along with their PubChem ids. Compound 1-20 (left to right) are:** *Apigenin - 5280443, Arbutin - 440936, Aromadendrin - 122850, Catechin - 9064, Chrysin - 5281607, Cirsiliol - 160237, Cirsimaritin - 188323, Desmethoxycentauridin - 5469524, Diosmetin - 5281612, Eriodictyol - 440735, Genestein - 5280961, Isovitexin - 162350, Kaempherol - 5280863, Luteolin - 5280445, Quercetin - 5280343, Sakuranetin -73571, Salvigenin - 161271, Sorbifolin - 3084390, Taxifolin - 439533, Xanthomicrol – 73207*.

**3.2. Retrieval of Receptor**

NS2B-NS3 protease (PDB id: 5GXJ) serves as an effective drug target in case of Zika infection due to its prominent role in the entry of the virus and replication within the host cell. So it was selected as the target protein molecule against the ligands in this study. The 3D structure of the molecule on the Protein Data Bank was directly downloaded from the PDB database, as shown in Fig.3.

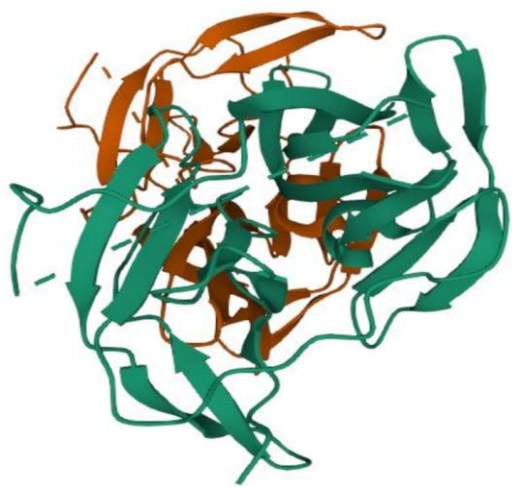

**Fig.3: The NS2B-NS3 protease (5gxj) is an effective drug target as it assists the virus to enter into the host body and replicating its genomic materials within the host cell. The 3D structure was extracted from the PBD database.**

**Active site prediction**

CASTp 3.0 has been used to predict the active binding site present in our protein target. Thirty-one active binding sites were identified, out of which the best binding pocket was selected. The area (SA) and the volume (SA) of the best binding pocket were found to be 995.342 and 617.549, respectively, as shown in Fig.4.

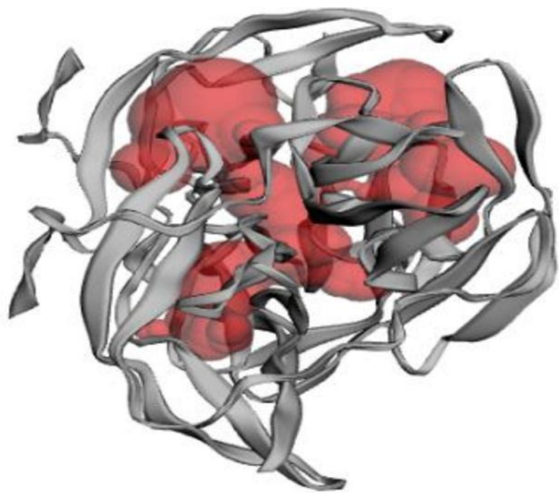

**Fig 4: The best binding pocket out of the 31 sites was selected, and the 3D structure along with the area (SA), i.e., 995.342, and volume (SA), i.e., 617.549, was extracted from CASTp.**

**ADMET analysis**

All the 20 selected flavonoids were screened using SWISS ADME, a drug screening tool. Ten compounds were screened at a time, and their canonical SMILES were obtained from the PubChem database. We set several parameters for scrutinizing the drug based on their physicochemical properties (see Table 1), Lipophilicity analysis (see Table 2), and drug-likeness analysis (see Table 3) based on the rules of the Lipinski filter, Ghose filter, Veber filter, Egan filter, and Muegge filter as well as bioavailability scores. Just high biological activities and low toxicity levels are not enough to make a compound satisfactorily accepted drug candidate. Hence, it is necessary to examine the ADMET profile. Lipinski's rule of five is one of the essential standards to screen the compound if it is orally consumable. It suggests that the compounds with the best properties to be considered an effective drug candidate should have the subsequent regulations: the molecular weight should be ≤ 500 DA, the number of hydrogen bond donors should be ≤ 5, hydrogen bond acceptors should be ≤ 10, and MLogP should be ≤ 5. The compound with the least number of violations will be considered an effective drug candidate. The Ghose filter determines the drug-likeness of the compound based on specific parameters, which state that the WLogP value should range between -0.4 and 5.6, the molecular weight should be between 160 and 480, a molecular refractivity (MR) should be between 40 and 130, and the whole number of atoms must 20 to 70. Veber's rule regulates the vital standards that should be fulfilled for a compound to be acceptable for oral bioavailability. The rules state that the Topological Polar Surface Area (TPSA) should be ≤ 140Å2, and the number of rotatable bonds should be ≤ 10. Egan's rule states that a molecule has good oral bioavailability if the WLogP value is ≤ 5.88 and the TPSA value ≤ 131.6Å. Muegge's rule states that a molecule is a suitable drug if the molecular weight lies between 200 and 600, XLogP lies between -2 and 5, the TPSA should be ≤ 150Å, the number of H-bond donors and acceptors should be ≤ 5 and 10 respectively. Based on (Table 1), (Table 2), and (Table 3) depicts the violations of the filters specified, which are carried out by the selected molecules.

**Table 1: Physicochemical Analysis of the shortlisted compounds using SWISS ADME.**

| Ligands | Molecular weight (g/mol) | Number of heavy atoms | Number of rotatable bonds | H-bond acceptor | H-bond donor | Molar refractivity | TPSA (Å$^2$) |
|---|---|---|---|---|---|---|---|
| *Apigenin* | 270.24 | 20 | 1 | 5 | 3 | 73.99 | 90.90 |
| *Arbutin* | 272.25 | 19 | 3 | 7 | 5 | 62.61 | 119.61 |
| *Aromadendrin* | 288.25 | 21 | 1 | 6 | 4 | 72.73 | 107.22 |
| *Catechin* | 290.27 | 21 | 1 | 6 | 5 | 74.33 | 110.38 |
| *Chrysin* | 254.24 | 19 | 1 | 4 | 2 | 71.97 | 70.67 |
| *Cirsiliol* | 330.29 | 24 | 3 | 7 | 3 | 86.97 | 109.36 |
| *Cirsimartin* | 314.29 | 23 | 3 | 6 | 2 | 84.95 | 89.13 |
| *Desmethoxycenturadin* | 330.29 | 24 | 3 | 7 | 3 | 86.97 | 109.36 |
| *Diosmetin* | 300.26 | 22 | 2 | 6 | 3 | 80.48 | 100.13 |
| *Eriodictyol* | 288.25 | 21 | 1 | 6 | 4 | 73.59 | 90.90 |
| *Genistein* | 270.24 | 20 | 1 | 5 | 3 | 73.99 | 234.29 |
| *Isovitexin* | 432.38 | 31 | 3 | 10 | 7 | 106.61 | 181.05 |
| *Kaempferol* | 286.24 | 21 | 1 | 6 | 4 | 76.01 | 111.13 |
| *Luteolin* | 286.24 | 21 | 1 | 6 | 4 | 76.01 | 190.28 |
| *Quercetin* | 302.24 | 22 | 1 | 7 | 5 | 78.03 | 131.36 |
| *Sakuranetin* | 286.28 | 21 | 2 | 5 | 2 | 76.04 | 75.99 |
| *Salvigenin* | 328.32 | 24 | 4 | 6 | 1 | 89.42 | 78.13 |
| *Sorbifolin* | 300.26 | 22 | 2 | 6 | 3 | 80.48 | 100.13 |
| *Taxifolin* | 304.25 | 22 | 1 | 7 | 5 | 74.76 | 127.45 |
| *Xanthomicrol* | 344.32 | 25 | 4 | 7 | 2 | 91.44 | 98.36 |

**Table 2: Lipophilicity analysis.**

| Ligands | XLogP | WLogP | MLogP | cLogP |
|---|---|---|---|---|
| *Apigenin* | 3.02 | 2.58 | 0.52 | 2.11 |
| *Arbutin* | -1.35 | -1.43 | -1.49 | -0.88 |
| *Aromadendrin* | 1.31 | 1.16 | 0.10 | 0.99 |
| *Catechin* | 0.36 | 1.22 | 0.24 | 0.85 |
| *Chrysin* | 3.52 | 2.87 | 1.08 | 2.55 |
| *Cirsiliol* | 3.07 | 2.59 | -0.07 | 2.13 |
| *Cirsimartin* | 3.32 | 2.89 | 0.47 | 2.46 |
| *Desmethoxycenturadin* | 3.07 | 2.59 | -0.07 | 2.13 |
| *Diosmetin* | 3.10 | 2.59 | 0.22 | 2.19 |
| *Eriodictyol* | 2.02 | 1.89 | 0.16 | 1.45 |
| *Genistein* | 2.67 | 2.58 | 0.52 | 2.04 |
| *Isovitexin* | 0.21 | -0.23 | -2.02 | 0.05 |
| *Kaempferol* | 1.90 | 2.28 | -0.03 | 1.58 |
| *Luteolin* | 2.53 | 2.28 | -0.23 | 1.73 |
| *Quercetin* | 1.54 | 1.99 | -0.56 | 1.23 |
| *Sakuranetin* | 2.85 | 2.49 | 0.96 | 2.25 |
| *Salvigenin* | 3.64 | 3.19 | 0.70 | 2.88 |
| *Sorbifolin* | 2.99 | 2.59 | 0.22 | 2.11 |
| *Taxifolin* | 0.95 | 0.86 | -0.64 | 0.63 |
| *Xanthomicrol* | 2.94 | 2.90 | 0.17 | 2.42 |

**Table 3: Drug-likeness analysis on the basis of Lipinski, Ghose, Veber, Egan, and Muegge.**

| Ligands | Lipinski filter | Ghose filter | Veber filter | Egan filter | Muegge filter | Bioavailability Score |
|---|---|---|---|---|---|---|
| *Apigenin* | Yes | Yes | Yes | Yes | Yes | 0.55 |
| *Arbutin* | Yes | Yes | No | Yes | Yes | 0.55 |
| *Aromadendrin* | Yes | Yes | Yes | Yes | Yes | 0.55 |
| *Catechin* | Yes | Yes | Yes | Yes | Yes | 0.55 |
| *Chrysin* | Yes | Yes | Yes | Yes | Yes | 0.55 |
| *Cirsiliol* | Yes | Yes | Yes | Yes | Yes | 0.55 |
| *Cirsimartin* | Yes | Yes | Yes | Yes | Yes | 0.55 |
| *Desmethoxycenturadin* | Yes | Yes | Yes | Yes | Yes | 0.55 |
| *Diosmetin* | Yes | Yes | Yes | Yes | Yes | 0.55 |
| *Eriodictyol* | Yes | Yes | Yes | Yes | Yes | 0.55 |
| *Genistein* | Yes | Yes | Yes | Yes | Yes | 0.55 |
| *Isovitexin* | Yes | Yes | No | No | No | 0.55 |
| *Kaempferol* | Yes | Yes | Yes | Yes | Yes | 0.55 |
| *Luteolin* | Yes | Yes | Yes | Yes | Yes | 0.55 |
| *Quercetin* | Yes | Yes | Yes | Yes | Yes | 0.55 |
| *Sakuranetin* | Yes | Yes | Yes | Yes | Yes | 0.55 |
| *Salvigenin* | Yes | Yes | Yes | Yes | Yes | 0.55 |
| *Sorbifolin* | Yes | Yes | Yes | Yes | Yes | 0.55 |
| *Taxifolin* | Yes | Yes | Yes | Yes | Yes | 0.55 |
| *Xanthomicrol* | Yes | Yes | Yes | Yes | Yes | 0.55 |

**BOILED-Egg**

The SWISS ADME feature, BOILED- Egg, was used to determine the parameters like gastrointestinal absorption, blood-brain barrier, and permeability glycoprotein substrate of the selected 20 ligands to check their likability to being accepted as a good and effective drug candidate. Ten compounds were analyzed at a time, as shown in **Fig.5A**, followed by the other 10, as shown in **Fig.5B**. Based on previous studies, we inferred that the molecules lying in the white region showed high GI absorption, and molecules lying in the yellow region depicted a high Blood-brain barrier **(see Table 4)**.

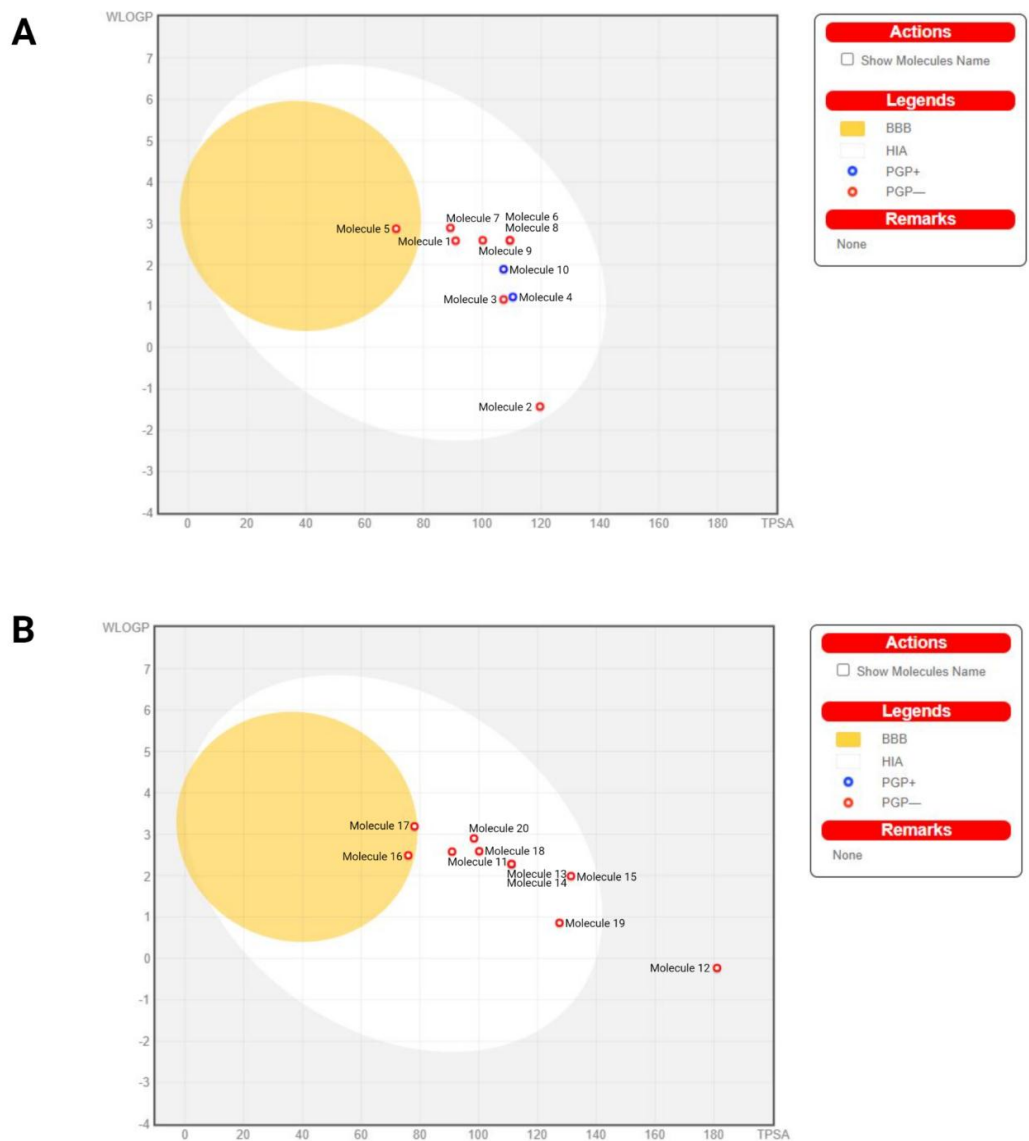

**Fig.5 BOILED-Egg analysis.**

A - Molecules 1-10 are depicted here in the following sequence: Apigenin, Arbutin, Aromadendrin, Catechin, Chrysin, Cirsiliol, Cirsimaritin, Desmethoxycentauridin, Diosmetin, Eriodictyol. The above BOILED-Egg analysis found that Chrysin lying in the yellow region had a high Blood-Brain Barrier, and the rest showed high GI absorption. Catechin and Eriodictoyl were positive for PGP substrate.

**B** - Molecules 11-20 are depicted here in the following sequence: Genistein, Isovitexin, Kaempferol, Luteolin, Quercetin, Sakuranetin, Salvigenin, Sorbifolin, Taxifolin, Xanthomicrol. From the above BOILED-Egg analysis, it was found that Sakuranetin and Salvigenin lying in the yellow region had a high Blood-Brain Barrier, and the rest showed high GI absorption except for Isovitexin since it did not lie within the white region.

**Table 4: Pharmacokinetics Analysis.**

| Ligands | Blood brain barrier | GI absorption | Permeability glycoprotein substrate |
|---|---|---|---|
| *Apigenin* | No | High | No |
| *Arbutin* | No | High | No |
| *Aromadendrin* | No | High | No |
| *Catechin* | No | High | Yes |
| *Chrysin* | Yes | High | No |
| *Cirsiliol* | No | High | No |
| *Cirsimartin* | No | High | No |
| *Desmethoxycenturadin* | No | High | No |
| *Diosmetin* | No | High | No |
| *Eriodictyol* | No | High | Yes |
| *Genistein* | No | High | No |
| *Isovitexin* | No | Low | No |
| *Kaempferol* | No | High | No |
| *Luteolin* | No | High | No |
| *Quercetin* | No | High | No |
| *Sakuranetin* | Yes | High | No |
| *Salvigenin* | Yes | High | No |
| *Sorbifolin* | No | High | No |
| *Taxifolin* | No | High | No |
| *Xanthomicrol* | No | High | No |

**Molecular docking analysis**

After the docking, all the selected flavonoids' binding affinities and docking scores were analyzed simultaneously (Table 5). A more negative value for the binding affinity suggests a better orientation of the ligands in the binding sites. Based on the binding affinity, docking scores, and the parameters stated in the above drug-likeness analysis, four compounds showing higher binding affinity and least violations of the drug-likability filter were further shortlisted from the 20 compounds. These four compounds were Quercetin, Isovitexin, Cirsiliol, and Salvigenin.

Further, these four compounds were analyzed for the target protein receptor NS2B-NS3 protease using Biovia Discovery Studio software to examine their interaction with the receptor. We used the data to determine the best interactions, as shown in Fig.6. Isovitexin (Fig.6A) showed a hydrogen bond interaction at the SER1033 residue of the NS2B-NS3 protease. In addition, it showed one pi-sigma interaction at THR1034, two pi-alkyl interactions at ALA1132, and one pi-alkyl interaction at VAL1036. Cirsiliol (Fig.6B) showed a hydrogen bond interaction at GLY1133 residue of the NS2B-NS3 protease. In addition, it showed one alkyl interaction at PRO1102, one pi-sigma interaction at THR1034, two pi-alkyl interactions at ALA1132, and one pi-alkyl interaction at VAL1036. Quercetin (Fig.6C) showed a hydrogen bond interaction at GLY1133 residue of the NS2B-NS3 protease. In addition, one pi-sigma interaction at THR1034 and one pi-alkyl interaction each at ALA1132 and VAL1036. Salvigenin Fig.6D) did not show any hydrogen bond interaction with the NS2B-NS3 protease. It showed one pi-sigma interaction at THR1034, one alkyl interaction at HIS1051, two pi-alkyl interactions at ALA1132, and one pi-alkyl interaction at VAL1036.

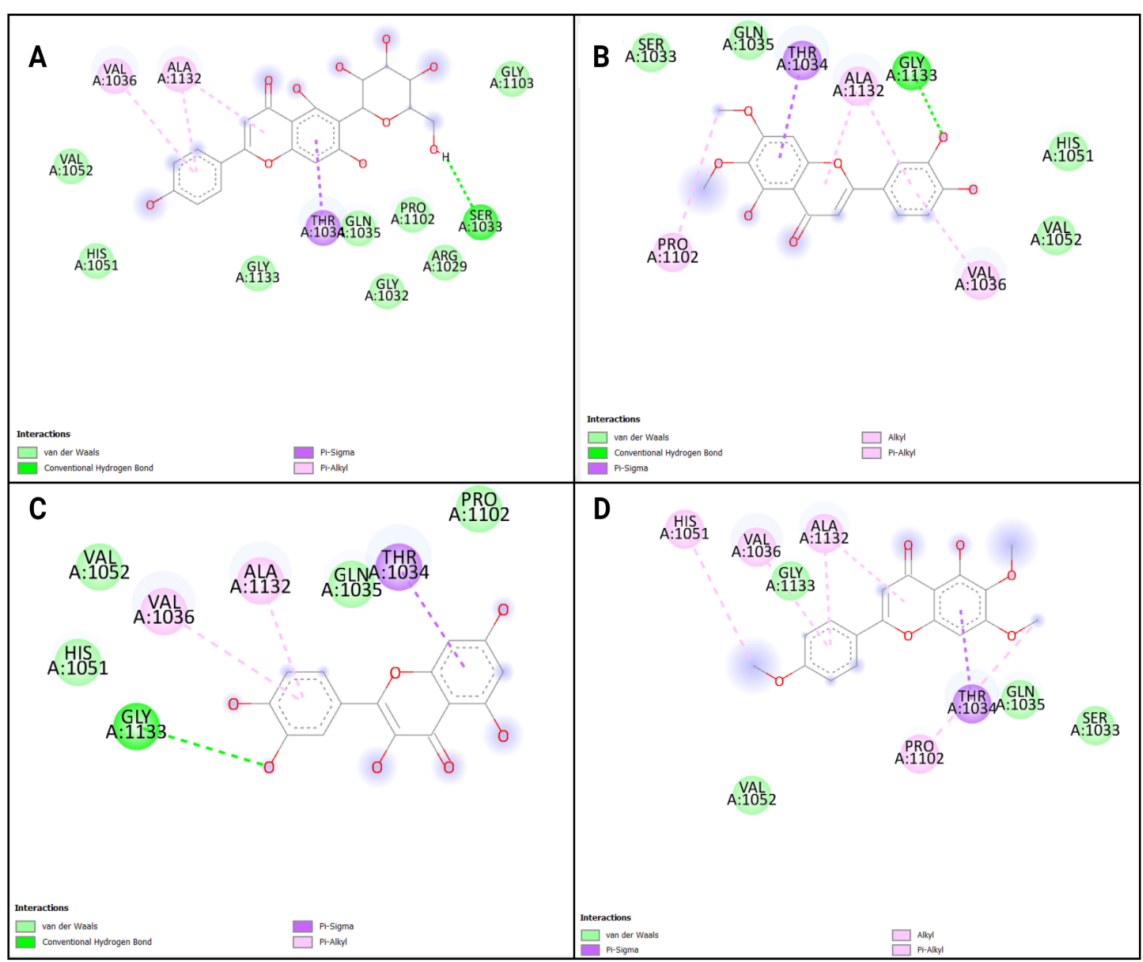

**Fig.6 Biovia Discovery Studio structural analysis.**

**Interaction of the receptor NS2B-NS3 protease with the ligands: A - Isovitexin, B - Cirsiliol, C - Quercetin, and D - Salvigenin, showing the 2D interactions of the ligands with the protein target.**

**Table 5: Binding affinity and docking scores.**

| Ligands | Binding Affinity (kcal/mol) | Docking scores |
|---|---|---|
| *Apigenin* | -8.1 | 4680 |
| *Arbutin* | -6.7 | 4362 |
| *Aromadendrin* | -7.5 | 4696 |
| *Catechin* | -7.7 | 4786 |
| *Chrysin* | -8.0 | 4414 |
| *Cirsiliol* | -8.5 | 5088 |
| *Cirsimartin* | -8.3 | 4934 |
| *Desmethoxycenturadin* | -8.2 | 4992 |
| *Diosmetin* | -8.1 | 4900 |
| *Eriodictyol* | -8.2 | 4674 |
| *Genistein* | -7.8 | 4656 |
| *Isovitexin* | -8.8 | 5428 |
| *Kaempferol* | -8.1 | 4838 |
| *Luteolin* | -8.4 | 4456 |
| *Quercetin* | -8.5 | 4712 |
| *Sakuranetin* | -7.9 | 4826 |
| *Salvigenin* | -8.4 | 5158 |
| *Sorbifolin* | -8.1 | 4766 |
| *Taxifolin* | -7.9 | 4736 |
| *Xanthomicrol* | -7.3 | 4720 |

**Molecular Dynamics (MD) studies**

Cirsiliol was chosen as the best hit and subjected to MD simulation testing. For MD simulation, the docked complex of Cirsiliol with the receptor NS2B-NS3 Protease was used. Normal mode analysis mobility allows us to examine the large-scale B-factor and mobility, as well as the molecule's stability [Fig.7]. The IMOD server made internal coordinates analysis available based on protein-ligand structural interactions. IMODs also determine the eigenvalue and measure the B-factor and structural deformity. The Normal Mode Analysis of the docked complex of our protein and hit ligand is shown in Fig.7. The deformability graph is depicted in Fig.7A The deformity graph depicted peaks in the graph that represented deformable regions in the protein. The B-Factor graph is shown in Fig.7B. The B-Factor, or main-chain deformability, is a measure of a molecule's ability to deform at each of its residues. Fig.7C depicts the complex's eigenvalue. The eigenvalue associated with each normal mode represents the motion stiffness. Its value is proportional to how much energy is required to distort the structure. The lower the eigenvalue, the simpler the deformation. Our docked complex had an eigenvalue of 2.630496e-05, indicating that our protein-ligand complex is easily deformed. The variance plot is depicted in Fig.7D. Individual variances are shown in red on the variance plot, while cumulative variance is shown in green. The covariance map is depicted in Fig.7E. This map depicts the correlation motion between two residues in red, the uncorrelated motion in white, and the anti-correlated motion in blue. Fig.7F depicts our docked complex's elastic map. Each dot in the graph represents one spring within the pair of atoms. The dots are colored based on stiffness, with darker grey dots representing stiffer springs and lighter grey dots representing softer springs. The molecular dynamics analysis revealed that our complex had a high degree of deformability. It also had a low eigenvalue, indicating that it could be easily deformed. The variance map showed more cumulative variances than an individual variance. The elastic network map produced similar results.

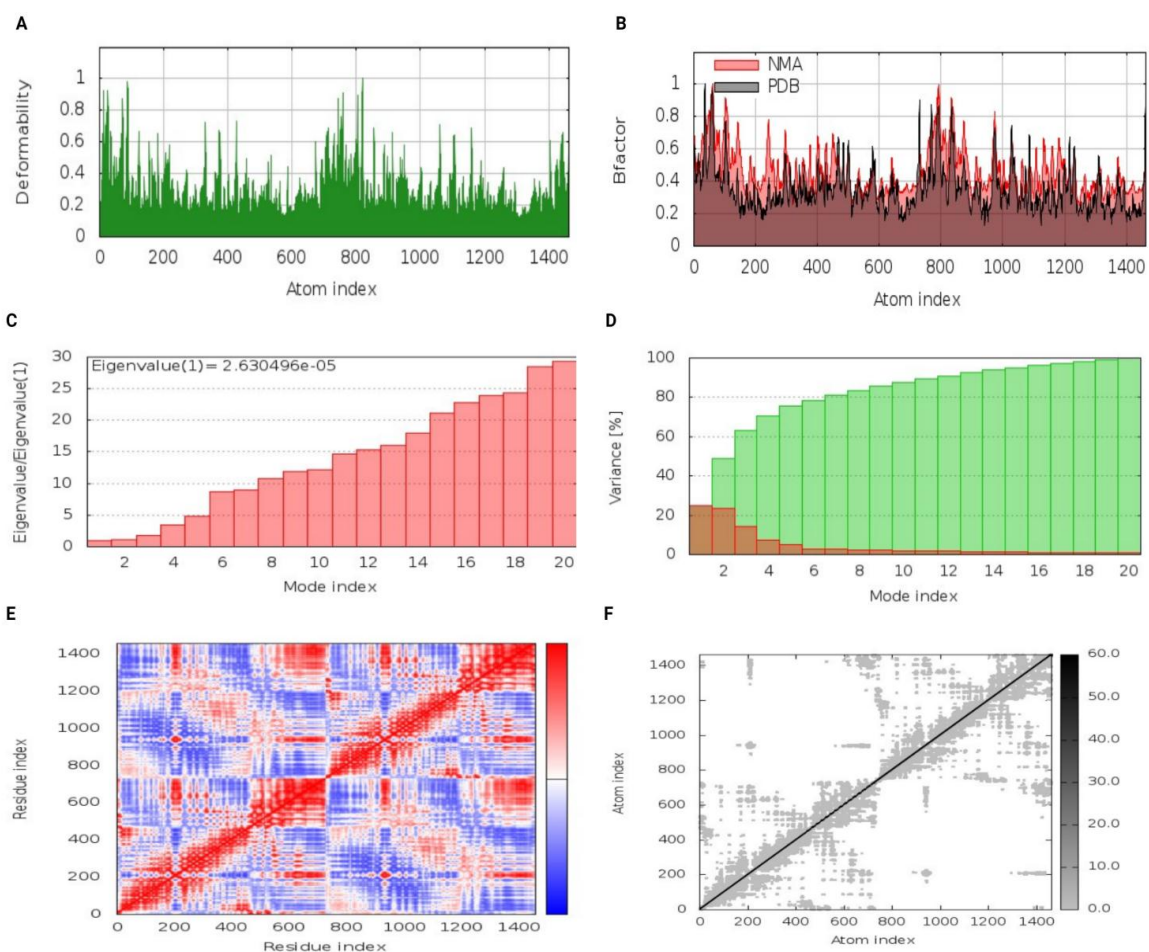

**Fig.7 Normal Mode Analysis of Cirsiliol with target receptor NS2B-NS3 Protease using iMODS software.**

For MD stimulation of our hit ligand-receptor complex using Desmond, we analyzed the RMSD of our ligand and protein receptor. The protein RMSD exhibited a stable trajectory throughout whereas the ligand RMSD was seen to show minor fluctuations between 40-80 ns but was stable for the rest of the process (Fig. 8). The Root Mean Square Fluctuation for our protein and ligand were analyzed further. As for the protein RMSF (P-RMSF), the highest fluctuation was observed at 3.8 Å (Fig.9). The ligand RMSF (L-RMSF) trajectory (Fig. 10) was found to be fairly stable. Apart from these, the interactions of the amino acids with the protein-ligand complex were also analyzed. From the protein-ligand interaction plot, it was seen that 26 amino acids were interacting with the protein-ligand complex (Fig.11). Among them, ALA_1087 and VAL_1146 showed the highest interactions which were mainly Hydrogen bonds and water bridges.

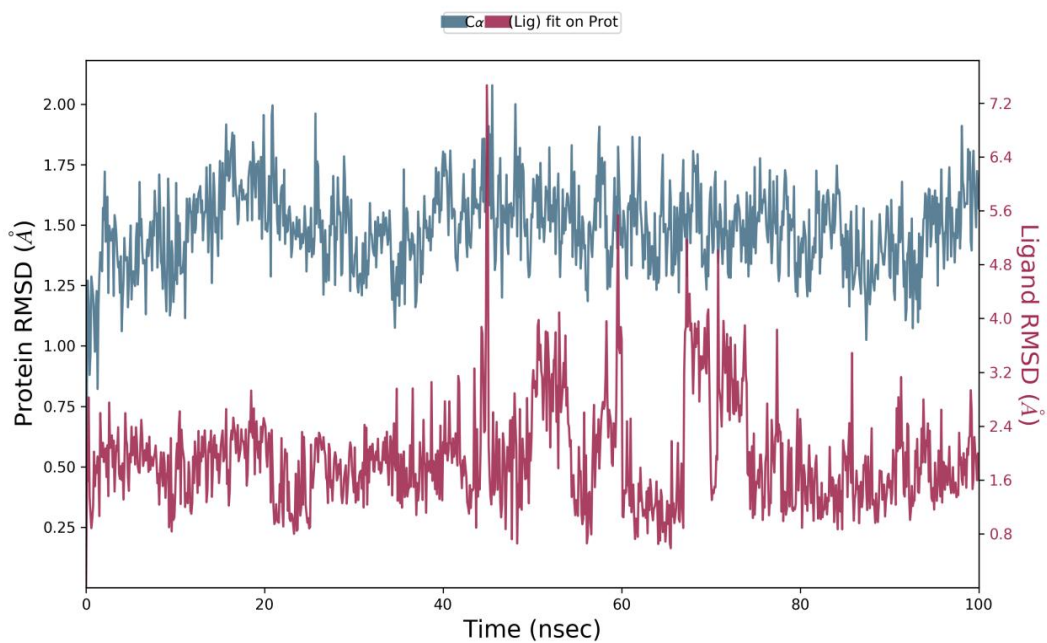

**Fig.8: Protein-Ligand RMSD: The Root Mean Square Deviation plot is used to calculate the average change in displacement of a set of atoms for a particular frame with respect to a reference frame. The plots above show the protein RMSD on the left Y-axis and the ligand RMSD on the right Y-axis. Initially all the protein frames aligned on the reference frame backbone, and then the protein RMSD is calculated based on the atom selection. We can get an insight into the structural conformation of the protein throughout the stimulation by monitoring its RMSD. A time frame of 100 ns of the Ligand RMSD indicates how stable the ligand Cirsiliol (PubChem ID: 160237) is with the protein NS2B-NS3 protease (PDB ID: 5gxj) and its binding pocket.**

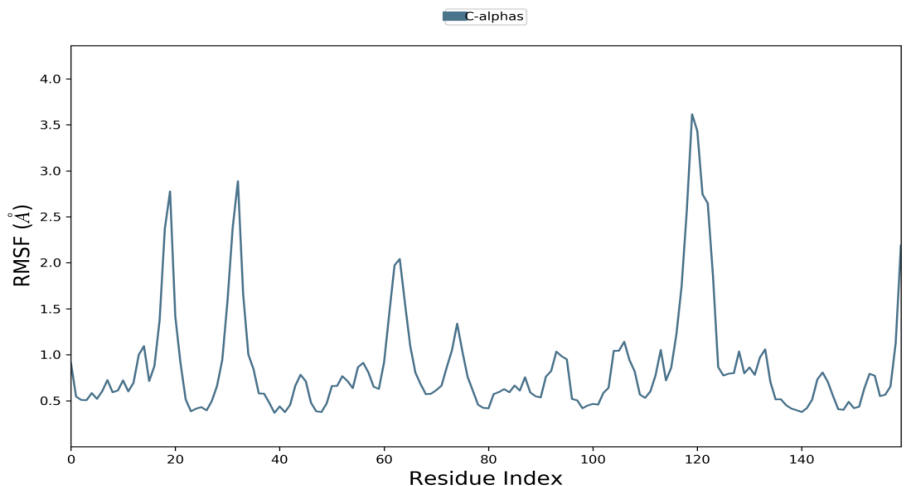

**Fig.9: Protein RMSF:** The protein RMSF is used for characterizing local changes along the protein chain. On this plot, peaks indicate areas of the protein that fluctuate the most during the simulation.

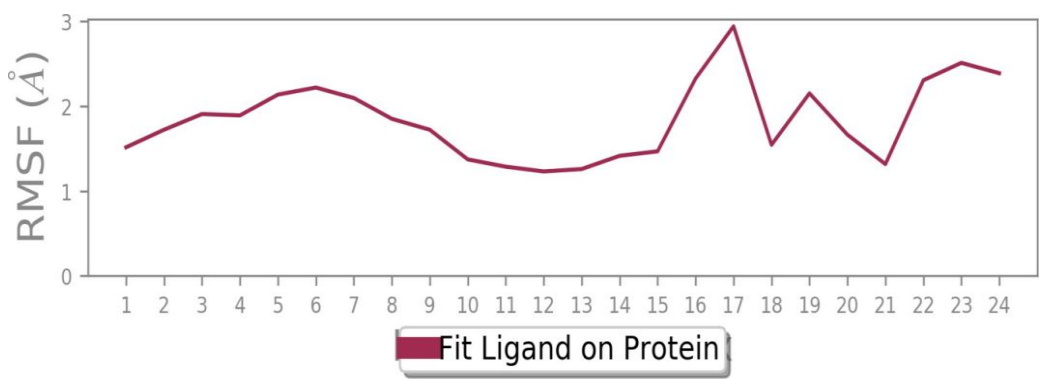

**Fig.10: Ligand RMSF:** The Ligand RMSF is used for characterizing the changes in ligand atom positions. This plot may give you an insight on how ligand fragments interact with the protein target and their entropic role in the binding event.

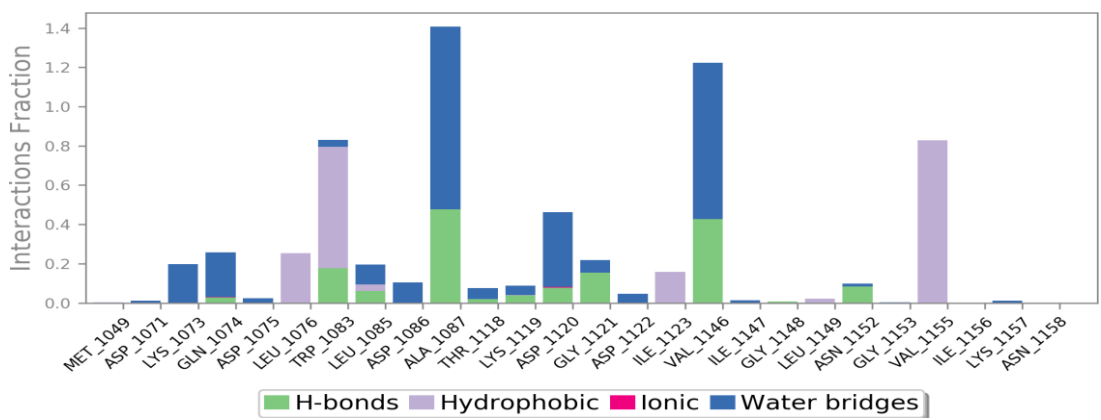

**Fig.11:** Protein-Ligand Contacts: Protein intearctions with the ligand can be monitored through the stimulation as shown in the above plots. Protein-ligand interactions can be categorized into four types: Hydrogen Bonds, Hydrophobic contacts, Ionic interactions and Water Bridges. Hydrogen Bonds (green) play an essential role in the ligand binding process. Hydrophobic contacts (violet) generally involve a hydrophobic amino acid on the protein target and an aromatic or aliphatic group on the ligand. Ionic interactions (pink) essentially occur between the two charged atoms that are closed to each other and are involved in hydrogen bond. Water Bridges (blue) are hydrogen bonded protein ligand interactions that are mediated by a water molecule. The X axis of this plot mentions all the amino acids of our protein target which are primarily involved in the formation of the protein ligand complex whereas the Y axis shows the interaction fraction of the different types of protein ligand interaction for each amino acid interacting with the ligand.

## 4. DISCUSSION

Oregano was and is still being used as a flavouring herb all over the world. However, it also shows optimistic results towards the treatment of many diseases, owing to its composition. Oregano traditionally has healing properties which have now dragged the attention of science for the betterment of humans. In our study, we have tried to explore the antiviral properties of the phytocompounds found in Oregano for treating Zika infection. The phytochemical compounds found in Oregano and selected for our research have shown good results with our target protein, the NS2B-NS3 Protease.

20 shortlisted flavonoids as mentioned in Table 1 were exposed to virtual screening using ADME analysis. Most of the phytocompounds in our study showed good Docking Scores and Binding affinity with our protein target, the NS2B-NS3 Protease. Isovitexin showed the best binding affinity of -8.8 kcal/mol followed by Quercetin (-8.5 kcal/mol), Cirsiliol (-8.5 kcal/mol) and Salvigenin (-8.4 kcal/mol). The ADME analysis also shows that almost all the compounds have a high GI absorption rate with Chrysin, Sakuranetin and Salvigenin having a very high Blood-Brain-Barrier permeation. Considering all the aspects such as Lipinski analysis, Binding Affinity, ADME Analysis and Ligand-Protein interactions on BIOVIA as shown in **(Fig.12)**, we selected Cirsiliol as our hit molecules. The second approach was to perform the molecular dynamics stimulations using Desmond to analyse the RMSD and RMSF of the ligand with respect to our target protein. This computer aided drug designing and proposed modelling were completely based on computational trials which can be used to find newer NS2B-NS3 protease inhibitors. The ligand-protein interactions of cirsiliol with NS2B-NS3 protease appeared to be fairly stable and hence cirsiliol can be further subjected to *in vitro* lab testing.

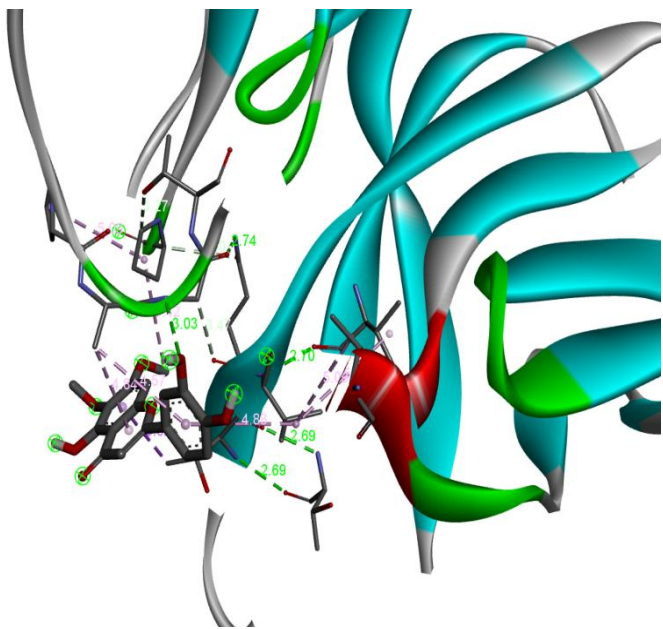

**Fig.12: Ligand-Protein 3D interaction: The best binding conformations and the corresponding interactions were determined. The protein backbone is displayed in solid ribbon style and the ligands with the active sites are displayed in stick representation. The distance between the ligand and protein is shown. The 3D interaction of the ligand Cirsiliol with the NS2B-NS3 protease is shown above.**

**CONCLUSION**

In this study, Cirsiliol and Isovitexin were found to be fulfilling all the parameters such as Lipinski analysis, ADMET analysis and binding affinity. Both the molecules were stable during the Molecular Dynamics analysis performed using Desmond at 100 ns. However, Cirsiliol showed better stability in comparison to Isovitein. Hence, our in-silico study indicates that the phytochemical compounds found in Oregano extensively pave the way to treating infection caused by Zika Virus.

**REFERENCES**

1. Kuno, G., Chang, G.-J. J., Tsuchiya, K. R., Karabatsos, N., & Cropp, C. B. (1998). Phylogeny of the Genus Flavivirus. Journal of Virology, 72(1): 73–83.
2. King AM, Lefkowitz E, Adams MJ, Carstens EB, editors. Virus taxonomy: ninth report of the International Committee on Taxonomy of Viruses. Elsevier, 2011 Nov 10.

3. Dick GW, Kitchen SF, Haddow AJ. Zika virus (II). Pathogenicity and physical properties. Transactions of the royal society of tropical medicine and hygiene, 1952; 46(5).

4. Who.int. n.d. Zika virus disease. [online] Available at: <https://www.who.int/health-topics/zika-virus-disease#tab=tab_1> [Accessed 29 December 2021].

5. Lancet T. Another kind of Zika public health emergency. Lancet (London, England), 2017 Feb 11; 389(10069): 573.

6. Lanciotti RS, Kosoy OL, Laven JJ, Velez JO, Lambert AJ, Johnson AJ, Stanfield SM, Duffy MR. Genetic and serologic properties of Zika virus associated with an epidemic, Yap State, Micronesia, 2007. Emerging infectious diseases, 2008 Aug; 14(8): 1232.

7. Sirohi D, Chen Z, Sun L, Klose T, Pierson TC, Rossmann MG, Kuhn RJ. The 3.8 Å resolution cryo-EM structure of Zika virus. Science, 2016 Apr 22; 352(6284): 467-70.

8. Sirohi D, Kuhn RJ. Zika virus structure, maturation, and receptors. The Journal of infectious diseases, 2017 Dec 15; 216(suppl_10): S935-44.

9. Bollati M, Alvarez K, Assenberg R, Baronti C, Canard B, Cook S, Coutard B, Decroly E, de Lamballerie X, Gould EA, Grard G. Structure and functionality in flavivirus NS-proteins: perspectives for drug design. Antiviral research, 2010 Aug 1; 87(2): 125-48.

10. Lee H, Ren J, Nocadello S, Rice AJ, Ojeda I, Light S, Minasov G, Vargas J, Nagarathnam D, Anderson WF, Johnson ME. Identification of novel small molecule inhibitors against NS2B/NS3 serine protease from Zika virus. Antiviral research, 2017 Mar 1; 139: 49-58.

11. Lee I, Bos S, Li G, Wang S, Gadea G, Desprès P, Zhao RY. Probing molecular insights into Zika virus–host interactions. Viruses, 2018 May; 10(5): 233.

12. Musso D, Bossin H, Mallet HP, Besnard M, Broult J, Baudouin L, Levi JE, Sabino EC, Ghawche F, Lanteri MC, Baud D. Zika virus in French Polynesia 2013–14: anatomy of a completed outbreak. The Lancet Infectious Diseases, 2018 May 1; 18(5): e172-82.

13. Swaminathan S, Schlaberg R, Lewis J, Hanson KE, Couturier MR. Fatal Zika virus infection with secondary nonsexual transmission. New England Journal of Medicine, 2016 Nov 10; 375(19): 1907-9.

14. Karimi O, Goorhuis A, Schinkel J, Codrington J, Vreden SG, Vermaat JS, Stijnis C, Grobusch MP. Thrombocytopenia and subcutaneous bleedings in a patient with Zika virus infection. The Lancet, 2016 Mar 5; 387(10022): 939-40.

15. Carteaux G, Maquart M, Bedet A, Contou D, Brugières P, Fourati S, Cleret de Langavant L, de Broucker T, Brun-Buisson C, Leparc-Goffart I, Mekontso Dessap A. Zika virus associated with meningoencephalitis. New England Journal of Medicine, 2016 Apr 21; 374(16): 1595-6.

16. Rasmussen SA, Jamieson DJ, Honein MA, Petersen LR. Zika virus and birth defects—reviewing the evidence for causality. New England Journal of Medicine, 2016 May 19; 374(20): 1981-7.

17. Dirlikov E, Torres JV, Martines RB, Reagan-Steiner S, Pérez GV, Rivera A, Major C, Matos D, Muñoz-Jordan J, Shieh WJ, Zaki SR. Postmortem findings in patient with Guillain-Barré syndrome and Zika virus infection. Emerging infectious diseases, 2018 Jan; 24(1): 114.

18. Sharma V, Sharma M, Dhull D, Sharma Y, Kaushik S, Kaushik S. Zika virus: an emerging challenge to public health worldwide. Canadian Journal of Microbiology, 2020; 66(2): 87-98.

19. Middleton E. Effect of plant flavonoids on immune and inflammatory cell function. Flavonoids in the living system, 1998; 175-82.

20. Lalani S, Poh CL. Flavonoids as antiviral agents for Enterovirus A71 (EV-A71). Viruses, 2020 Feb; 12(2): 184.

21. Jacob A, Thomas J. Therapeutic potential of dietary flavonoids against viral-borne infections: A review. Drug Invention Today, 2019 Feb 1; 11(2).

22. Elfiky, A. A., & Ismail, A. M. (2018). Molecular docking revealed the binding of nucleotide/side inhibitors to Zika viral polymerase solved structures. *SAR and QSAR in Environmental Research*, *29*(5): 409-418.

23. Bhargava, S., Patel, T., Gaikwad, R., Patil, U. K., & Gayen, S. (2019). Identification of structural requirements and prediction of inhibitory activity of natural flavonoids against Zika virus through molecular docking and Monte Carlo based QSAR Simulation. *Natural product research*, *33*(6): 851-857.

24. Sangeetha, K., Martín-Acebes, M. A., Saiz, J. C., & Meena, K. S. (2020). Molecular docking and antiviral activities of plant derived compounds against zika virus. *Microbial Pathogenesis*, *149*: 104540.

25. Rasool, N., Jalal, A., Amjad, A., & Hussain, W. (2018). Probing the pharmacological parameters, molecular docking and quantum computations of plant derived compounds exhibiting strong inhibitory potential against NS5 from Zika virus. *Brazilian Archives of Biology and Technology*, *61*.

26. Proestos C, Komaitis M. Analysis of naturally occurring phenolic compounds in aromatic plants by RP-HPLC coupled to diode array detector (DAD) and GC-MS after silylation. Foods, 2013 Mar; 2(1): 90-9.

27. Vujicic M, Nikolic I, Kontogianni VG, Saksida T, Charisiadis P, Orescanin-Dusic Z, Blagojevic D, Stosic-Grujicic S, Tzakos AG, Stojanovic I. Methanolic extract of Origanum vulgare ameliorates type 1 diabetes through antioxidant, anti-inflammatory and anti-apoptotic activity. British Journal of Nutrition, 2015 Mar; 113(5): 770-82.

28. Hawas UW, El-Desoky SK, Kawashty SA, Sharaf M. Two new flavonoids from Origanum vulgare. Natural Product Research, 2008 Nov 20; 22(17): 1540-3.

29. Tuttolomondo T, La Bella S, Licata M, Virga G, Leto C, Saija A, Trombetta D, Tomaino A, Speciale A, Napoli EM, Siracusa L. Biomolecular characterization of wild sicilian oregano: Phytochemical screening of essential oils and extracts, and evaluation of their antioxidant activities. Chemistry & biodiversity, 2013 Mar; 10(3): 411-33.

30. Gonçalves S, Moreira E, Grosso C, Andrade PB, Valentão P, Romano A. Phenolic profile, antioxidant activity and enzyme inhibitory activities of extracts from aromatic plants used in Mediterranean diet. Journal of food science and technology, 2017 Jan; 54(1): 219-27.

31. Hossain MB, Camphuis G, Aguiló- Aguayo I, Gangopadhyay N, Rai DK. Antioxidant activity guided separation of major polyphenols of marjoram (Origanum majorana L.) using flash chromatography and their identification by liquid chromatography coupled with electrospray ionization tandem mass spectrometry. Journal of separation science, 2014 Nov; 37(22): 3205-13.

32. Kim S, Chen J, Cheng T, Gindulyte A, He J, He S, Li Q, Shoemaker BA, Thiessen PA, Yu B, Zaslavsky L. PubChem in 2021: new data content and improved web interfaces. Nucleic acids research, 2021 Jan 8; 49(D1): D1388-95.

33. Berman HM, Westbrook J, Feng Z, Gilliland G, Bhat TN, Weissig H, Shindyalov IN, Bourne PE. The protein data bank. Nucleic acids research, 2000 Jan 1; 28(1): 235-42.

34. Tian W, Chen C, Lei X, Zhao J, Liang J. CASTp 3.0: computed atlas of surface topography of proteins. Nucleic acids research, 2018 Jul 2; 46(W1): W363-7.

35. Daina A, Michielin O, Zoete V. SwissADME: a free web tool to evaluate pharmacokinetics, drug-likeness and medicinal chemistry friendliness of small molecules. Scientific reports, 2017 Mar 3; 7(1): 1-3.

36. Lipinski CA, Lombardo F, Dominy BW, Feeney PJ. Experimental and computational approaches to estimate solubility and permeability in drug discovery and development settings. Advanced drug delivery reviews, 1997 Jan 15; 23(1-3): 3-25.

37. Ghose AK, Viswanadhan VN, Wendoloski JJ. A knowledge-based approach in designing combinatorial or medicinal chemistry libraries for drug discovery. 1. A qualitative and quantitative characterization of known drug databases. Journal of combinatorial chemistry, 1999 Jan 12; 1(1): 55-68.

38. Veber DF, Johnson SR, Cheng HY, Smith BR, Ward KW, Kopple KD. Molecular properties that influence the oral bioavailability of drug candidates. Journal of medicinal chemistry, 2002 Jun 6; 45(12): 2615-23.

39. Egan WJ, Merz KM, Baldwin JJ. Prediction of drug absorption using multivariate statistics. Journal of medicinal chemistry, 2000 Oct 19; 43(21): 3867-77.

40. Muegge I. Pharmacophore features of potential drugs. Chemistry–A European Journal, 2002 May 3; 8(9): 1976-81.

41. Martin YC. A bioavailability score. Journal of medicinal chemistry, 2005 May 5; 48(9): 3164-70.

42. Daina A, Zoete V. A boiled- egg to predict gastrointestinal absorption and brain penetration of small molecules. ChemMedChem, 2016 Jun 6; 11(11): 1117-21.

43. Dallakyan S, Olson AJ. Small-molecule library screening by docking with PyRx. InChemical biology 2015 (pp. 243-250). Humana Press, New York, NY.

44. Schneidman-Duhovny D, Inbar Y, Nussinov R, Wolfson HJ. PatchDock and SymmDock: servers for rigid and symmetric docking. Nucleic acids research, 2005 Jul 1; 33(suppl_2): W363-7.

45. Biovia DS, Dsme R. San Diego: Dassault Systèmes. Release, 2015; 4.